%% file: we.tex
\documentclass[conference]{IEEEtran}
\usepackage{amssymb,stmaryrd,amsmath,amsfonts,rotating,mathrsfs, graphics, graphicx}
\usepackage[noadjust]{cite}
\usepackage{color}
\usepackage{epic}
\RequirePackage{bbm}
\input{./definition}
\allowdisplaybreaks
\begin{document}
\title{Near concavity of the growth rate for coupled LDPC chains}
\author{\IEEEauthorblockN{S. Hamed Hassani and Nicolas Macris}
\IEEEauthorblockA{Ecole Polytecnique F\'ed\'erale de Lausanne\\ School of Computer
 \& Communication Sciences\\ LTHC-Station 14, Lausanne, CH-1015, Switzerland\\
\{seyedhamed.hassani, nicolas.macris\}@epfl.ch}
\and
\IEEEauthorblockN{Ryuhei Mori}
\IEEEauthorblockA{Kyoto University\\
Department of Systems Science, Graduate School of Informatics\\
Kyoto, 606-8501, Japan\\
rmori@sys.i.kyoto-u.ac.jp}}

\maketitle

\begin{abstract}
\boldmath
Convolutional Low-Density-Parity-Check (LDPC) ensembles have excellent performance. 
Their iterative threshold increases with their average degree, or with the size of the
coupling window in randomized constructions. In
the later case, as the window size grows, 
the Belief Propagation (BP) threshold attains the maximum-a-posteriori (MAP) threshold of the underlying ensemble. 
In this contribution we show that a similar phenomenon happens for the growth rate 
of coupled ensembles. Loosely speaking, we observe that as the coupling strength grows, the growth 
rate of the coupled ensemble 
comes close to the concave hull of the underlying ensemble's growth rate. For
ensembles randomly coupled across a window the growth rate actually tends to the concave hull of the 
underlying one as the window size increases. Our observations are supported by the calculations of 
the combinatorial
growth rate, and that of the growth rate derived from the replica method. 
The observed concavity is a general feature
of coupled mean field graphical models and is already present at the level of coupled Curie-Weiss models.
There, the canonical free energy of the coupled system tends 
to the concave hull of the underlying one.
As we explain, the behavior of the growth rate of coupled ensembles is exactly
analogous.
\end{abstract}

\IEEEpeerreviewmaketitle

\section{Introduction}\label{1}

Convolutional low-density parity-check (LDPC) codes where initially introduced by Felstroem and 
Zigangirov \cite{Felstrom-Zizangirov} and since then they have spurred a large body of work
 (see \cite{Engdahl-Zizangirov},
\cite{Engdahl-Lentmaier-Zizangirov}, \cite{Lentmaier-Trubachev-Zizangirov}, 
\cite{Tanner-Sridhara-Sridhrara-Fuja-Costello}, 
\cite{Sridharan-Lentmaier-Costello-Zizangirov}, \cite{Lentmaier-Sridharan-Zizangirov-Costello}, 
\cite{Lentmaier-Fettweis-Zizangirov-Costello},
\cite{Kudekar-Richardson-Urbanke}, 
\cite{Kudekar-Measson-Richardson-Urbanke}). 
The constructions of convolutional LDPC ensembles involve the appropriate coupling, 
of standard individual LDPC ensembles, into a one dimensional chain. 
The main observation is that as the strength of the coupling (e.g. the average degrees or the size of the window) increases,  the 
BP threshold of the coupled ensemble comes close to the MAP one of the underlying individual ensemble. 
This provides a new versatile way of approaching capacity.

In this contribution we focus on the growth rate of the relative Hamming distance spectrum for coupled ensembles of LDPC codes (we will call 
this simply the {\it growth rate} in what follows, see sec. \ref{notation} for definitions). 
We study 
this quantity for two kinds of ensembles, by combinatorial as well as replica methods. Both of these approaches have been already used 
for standard LDPC ensembles (see \cite{Litsyn-Sheleyev}, \cite{Litsyn-Sheleyev-2}, \cite{Miller-Burstein}, \cite{Mourik}, \cite{Di-Montanari-Urbanke}). 
The growth rate of a generic standard
LDPC ensemble is not a concave curve and displays a gap near $0$ weight, indicating the absence of code words of macroscopic weight near $0$.
Here we show that, as the coupling strength increases, the growth rate of the convolutional LDPC ensembles becomes
nearly identical to the {\it concave hull} of the growth rate of the individual ensemble. This phenomenon is illustrated in fig. \ref{all-36}.
We conjecture that, for the ensembles with a randomized coupling accross a window of size $w$, 
in the limit where the length of the chain $L\to +\infty$ (first) and $w\to+\infty$ (second) the growth rate 
really tends to the concave hull
of the individual system curve.

The threshold saturation phenomenon is a general feature of coupled chains of mean field graphical models. 
The same phenomenon happens in Curie-Weiss (CW) chains \cite{HNR}, and also
coupled constraint satisfaction models \cite{itw}. As explained in more detail in sec. \ref{analogies} the concavity of the growth rate for coupled
LDPC chains is the precise analog of the concavity of the {\it canonical free energy} of a coupled CW chain. This analogy
in fact constituted our initial intuition for the behavior of the growth rate of coupled LDPC models. 

\section{Models and Results}\label{2}

\subsection{Notations and Preliminaries}\label{notation}

For the underlying individual ensemble we take a regular 
$\mathrm{LDPC}(n,l,r)$ ensemble
with $m= \frac{l}{r}n$  check nodes of degree $r$ and 
$n$ variable nodes of degree $l$.  
We define the relative weight of a code-word $(x_1,...,x_n)$ as 
$\omega = 1-\frac{2}{n}\sum_{i=1}^N x_i$ and call $A_n(\omega)$
the number of code words of relative weight $\omega$. Note that here $\omega\in [-1,\cdots, 1]$ with $\omega=1$ corresponding to the 
all $0$ codeword. The growth rate 
is 
\begin{equation}\label{growth-rate-def}
G_n(\omega)= \frac{1}{n}\mathbb{E} [\ln A_n(\omega)]
\end{equation}
where the expectation is over the $\mathrm{LDPC}(n,l,r)$ ensemble.
Applying Jensen's inequality one obtains an upper bound $G_n(\omega) \leq G_{n, \mathrm{com}}(\omega)$, sometimes called the combinatorial growth rate,
\begin{equation}\label{growth-rate-com}
G_{n, \mathrm{com}}(\omega)= \frac{1}{n}\ln\mathbb{E}[A_n(\omega)]
\end{equation}
Here we are interested in the infinite length 
quantities 
$G(\omega)$ and $G_{\mathrm{com}}(\omega)$. 
For the regular ensemble considered here the infinite 
length limits\footnote{It is an open problem to prove the existence of the first limit.} are believed to be equal: indeed the 
combinatorial calculation of $G_{\mathrm{com}}(\omega)$ and the replica method calculation of $G(\omega)$ lead to the same 
set of equations (see \cite{Di-Montanari-Urbanke} and for a partial proof \cite{Macris}). 
One first fixes $\omega$ and solves for $(y,z,h)$ the 
set of equations
\begin{equation}
\begin{cases}
y & =z^{r-1}, 
\\
z & =\tanh(h+(l-1)\tanh^{-1}(y)),
\\
\omega & = \tanh (h + l\tanh^{-1} y). 
\end{cases}
\label{fixed-point}
\end{equation}
For $\omega=1$ there is the family of trivial solutions $(y=z=1, h\in \mathbb{R})$; for $\omega<1$ there exist a non-trivial
single valued solution $(y(\omega), z(\omega), h(\omega)$) from which one can compute  
\begin{align}
G(\omega) = & \frac{l}{r}\ln\bigl(\frac{1+ z^r}{2}\bigr)
+\ln\bigl(e^h(1+y)^l + e^{-h}(1-y)^l\bigr)  
\nonumber \\
&
 - l\ln(1+zy) -\omega h
\end{align}
Both $G(\omega)$ and $h(\omega)$ are illustrated as dotted lines on fig. \ref{wiggle} for a $(3,6)$ ensemble.
In the first three terms one finds the contributions of $n\frac{l}{r}$ checks, $n$ variable nodes, and $nl$ edges.
It is easily checked that the {\it total} derivative satisfies
\begin{equation}\label{derivative}
\frac{d G(\omega)}{d\omega} = - h(\omega).
\end{equation}
This formula is useful because it allows to reconstruct the growth rate by integrating 
$h(\omega)$ (note $G(1)=0$)
\begin{equation}\label{integral}
G(\omega) = \int_\omega^1 d\omega h(\omega).
\end{equation}
 
\subsection{The  coupled  ensembles}\label{coupled}

\indent{\it The $(l,r,L)$ ensemble.}
We assume that $\frac{r}{l}$ is integer.  
At each position $i \in
\{-L,\cdots, L\}$ of a one-dimensional chain (of length $2L+1$) we lay down sets of $m$ checks
and $n$ variable nodes. At the right boundary, we also add additional sets of $m$ check nodes  in positions $L+1, \cdots, L+l-1$.
Each check (resp. variable) in the range $[-L, \cdots, L]$ has a set of $r$ edges that we view as a $l$ groups of $\frac{r}{l}$ edges each.
Fix $i$. For each $k \in \{0, \cdots, l-1\}$, $n$ edges emanating from $n$ variable nodes at position $i$ 
are connected - through a uniformly random permutation - to  a group of $m\frac{r}{l}$ edges emanating the checks at position $i+k$.
The couple $(i, i+k)$ will parametrize the class of edges connecting variable nodes at position $i$ to checks at position $i+k$. 
\\
\indent{\it The $(l,r, w, L)$ ensemble.}
This ensemble is a modification of the previous one obtained by adding a randomization of 
the edge connections. We refer to \cite{Kudekar-Richardson-Urbanke} for the detailled construction.  
It is no longer assumed that $\frac{r}{l}$ is an integer, and there is a new parameter $w\in \mathbb{N}$, the window size.  
In the limit of large size, each edge of a variable node at position $i$ is uniformly 
and independently connected to a check in
the range $\{i - w +1,\cdots,  i \}$; and each check at position $i$ is uniformly and independently connected 
to a variable node in $\{i,\cdots, i - w +1\}$. It can be shown that 
for $n$ and $m$ large the connections of a variable nodes at position $i$ 
towards checks are  independ and uniform in the range 
$\{i, \cdots, i+w-1\}$ and that variable nodes have constant degree $l$. 

The relative weight $\omega$ of code words $(x_{j,i})_{j=1,i=-L}^{N,L}$ is 
\begin{equation}
\omega=\frac{1}{2L+1}\sum_{i=-L}^L \omega(i),\qquad \omega(z) = 1-\frac{2}{nL}\sum_{j=1}^{n} x_{j,i}
\end{equation}
and the growth is defined as in \eqref{growth-rate-def} but with the normalization $nL$ in front of the $log$. We denoted it by $G_1(\omega)$ 
for the $(l,r,L)$ ensemble and 
$G(\omega_2)$ for the $(l,r,w,L)$ ensemble. The corresponding combinatorial growth rates are $G_{1, \mathrm{com}}(\omega)$ 
and $G_{2, \mathrm{com}}(\omega)$.

\subsection{Main result: concavity of the growth rate for the coupled ensembles}

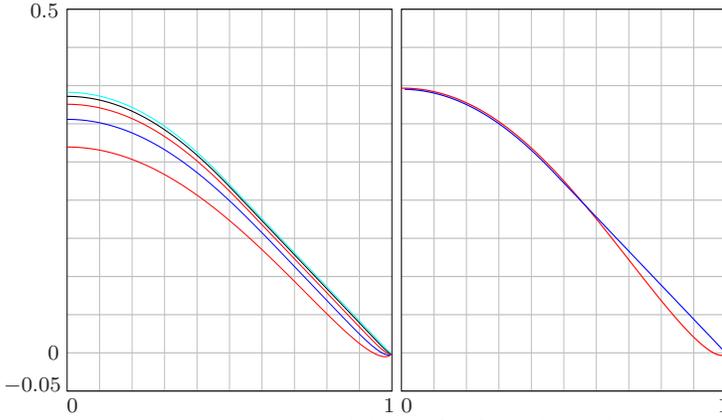
\begin{figure}[h!]
\begin{centering}
\input{all-36}
\caption{\small{Left: Evolution of $G_1(\omega)$, $\omega\in [0,1]$, for $(3,6,L)$ ensembles with $L=4, 8, 16, 32, 64$, 
Right: $(3,6,256)$ ensemble}
}
\label{all-36}
\end{centering}
\end{figure}

\indent{\it The $(l,r,L)$ ensemble.} 
As for a usual $(l,r)$ ensemble \cite{Di-Montanari-Urbanke}, in turns out that for the $(l,r,L)$ the replica and combinatorial approaches yield the same expressions
(up to boundary terms that vanish as $L\to +\infty$) for the growth rate. In sec. \ref{3} we sketch the combinatorial derivation which leads to a 
set of fixed point equations, which we have solved numericaly. Figure \ref{all-36} shows that the growth rate for the $(3,6,256)$ ensemble is indistinguishable from 
the concave hull of the underlying ensemble. Note that the picture is symmetric for $\omega$ negative because the check node degrees are even. Figure 
\ref{wiggle} shows both curves $G_1(\omega)$ and $h_1(\omega)$ for the $(3,6, 32)$ ensemble. It is clear that the 
concave hull of $G_1(\omega)$ develops hand in hand with the 
plateau in $h_1(\omega)$. The plateau is very close to a value $h_c=h(\omega_c)$ where $\omega_c$ is found from the concave hull construction
for $G(\omega)$ (i.e it is the tangency point
of a straight line going through $\omega=1$). Note that it can also be obtained directly from the $h(\omega)$ dotted curve by
a Maxwell equal area construction. 
A magnification of the curves shows 
that they display wigles of small amplitude. There are approximately $2L$ oscillations with a period $O(\frac{1}{L})$ that extend throughout the 
plateau of $h_1(\omega)$ and throughout the straight slope in $G_1(\omega)$. The amplitude of the wiggles in $G_1(\omega)$ is $\frac{1}{2L}$ times the amplitude 
of the wigles for $h_1(\omega)$ and therefore harder to detect. 
Fig. \ref{all-36} shows the sequence of curves as a function of $L$.
The limiting curve is numericaly very close to the concave hull of the underlying standard $(3,6)$ 
ensemble growth rate. Note however that there is no fundamental reason for it to be concave and this is not exactly the case. 
As explained below 
we conjecture that concavity is attained only in an appropriate limit with the $(l,r,w,L)$ ensemble. 

\indent{\it The $(l,r, w, L)$ ensemble.} 
In the randomized construction the combinatorial and replica expression are not the same, although they 
are numericaly close. It is generaly believed that the replica expressions are exact and we therefore stick those. 
In sec. \ref{4} we give the result of the replica calculation. 
The corresponding fixed point equations involve non-trivial densities and 
we have solved them by the method of population dynamics. The curves $h_2(\omega)$ and $G_2(\omega)$ will display 
the same features as in the previous ensemble so do not comment more on these. 
Our interest is in the behavior of the model as $w$ grows (when $L$ is formaly infinite).
For $h> h_{\mathrm{it}}^{(w,L)}$ there is
a unique trivial fixed point whereas for $h\leq h_{\mathrm{it}}^{(w,L)}$ two extra fixed points 
(besides the trivial one) appear. Table \ref{table} gives
the values of the bifurcation point for $w =1$ (uncoupled case), $w=2$ and $w=3$ when $L=20$. 
As $w$ grows $h_{\mathrm{it}}^{(w,L)} \to h_{c}$ and we conjecture
that $\lim_{w\to+\infty}\lim_{L\to +\infty} h_{\mathrm{it}}^{(w,L)} = h_c$. 
As a consequence, for the $(l,r, w, L)$ ensemble we conjecture that 
\begin{equation}
\lim_{w\to +\infty}\lim_{L\to +\infty}\lim_{n\to+\infty} \frac{1}{nL}\mathbb{E}[\ln A^{(2)}_n(\omega)] =  
\mathrm{concave~hull~} G(\omega)
\end{equation}
We stress that the order of limits matters.
Note that here we expect from the experience with the coupled CW chain 
(sec. \ref{analogies}) that the amplitude of the wiggles is exponentialy small in $w$ but this is difficult to observe.
 
\begin{figure}[h!]
\begin{centering}
\input{wiggle-we}
\caption{\small{ $G_1(\omega)$ for $\omega\in[0,1]$. 
Dotted curves: underlying $(3,6)$ ensemble. Full curves: $(3,6,32)$ ensemble }
}
\label{wiggle}
\end{centering}
\end{figure}
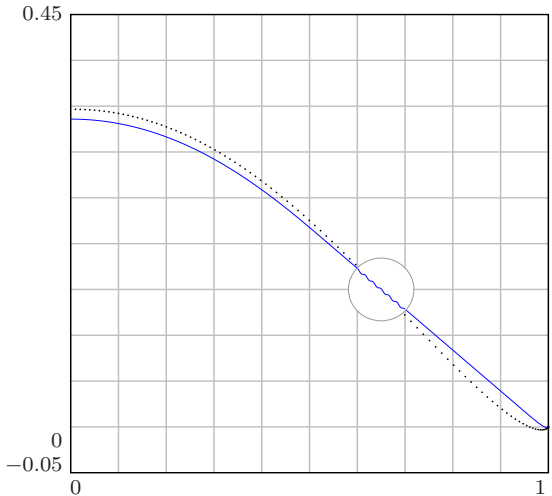

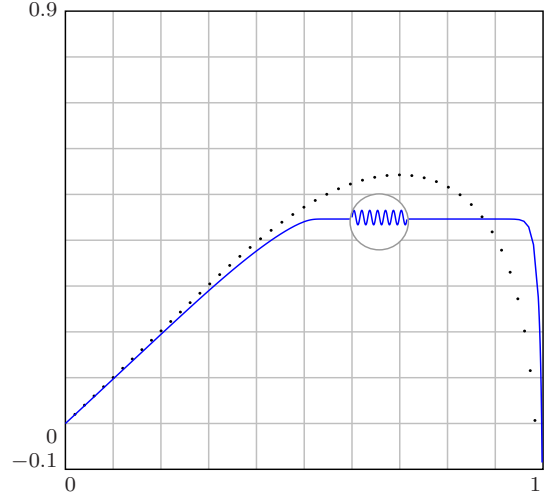
\begin{figure}[h!]
\begin{centering}
\input{wiggle-hw}
\caption{\small{$h_1(\omega)$ for for $\omega\in[0,1]$. 
Dotted curves: underlying $(3,6)$ ensemble. Full curves: $(3,6,32)$ ensemble }
}
\label{wiggle}
\end{centering}
\end{figure}
\begin{table}
\centering
\begin{tabular}{c c c c c c}
\hline\hline 
                         & $(3,6)$ & &   $(4,8)$  & &   $(5,10)$ \\
\hline
$h_c$                          &  $0.446$ & & $0.442 $ & &  $0.441$ \\ 
$h_{\text{it}}$              &  $0.543$ & & $0.629 $ & &  $0.706$ \\
$h_{\text{it}}^{2,20}$   &  $0.446$ & & $0.447 $ & &  $0.469$ \\
$h_{\text{it}}^{3,20}$   &  $0.446$ & & $0.442 $ & &  $0.442$ \\

\hline
\end{tabular}
\caption{{\small Thresholds of the 
underlying $(r,l)$ ensemble; and iterative thresholds of $(r,l,w, 20)$ ensembles}}
\label{table}
\end{table} 

%


\subsection{Analogies with the CW chain}\label{analogies}

\begin{figure}[h!]
\begin{centering}
\input{free-vdw}
\caption{\small{Dotted curves: free energy and van der Waals curve of the CW model for $J>1$ as functions of $m$.
Continuous curves: free energy and  
van-der-Waals isotherm of the coupled chain. 
Oscillations extend 
throughout the plateaus, have period $O(1/L)$  and amplitude exponentialy small in $w$.}
}
\label{vdw}
\end{centering}
\end{figure}
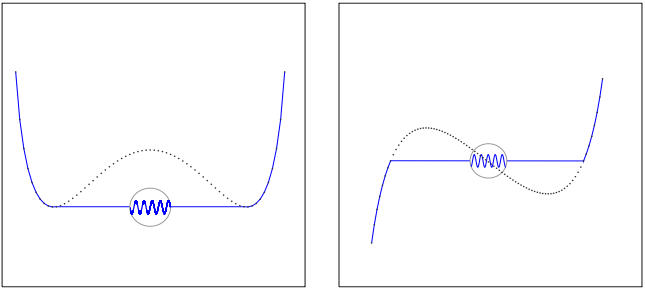

We briefly summarize the picture that arises in the simple setting of the coupled chain of Curie-Weiss models \cite{itw}, \cite{HNR}.
The Curie-Weiss 
model (CW) in the canonical ensemble (or lattice-gas interpretation) has Hamiltonian 
$
H_N = -\frac{J}{N}\sum_{\langle t,u\rangle} 
s_t s_u$
where the spins $s_t=\pm 1$ are attached to the $N$ vertices 
of a complete graph, the sum over $\langle t,u\rangle$ carries 
over all edges of the graph  and $J>0$. The free 
energy, for a fixed magnetization $m=\frac{1}{N}\sum_{t=1}^N s_t$, is easily computed
\begin{equation}\label{varia} 
\lim_{N\to+\infty}\frac{1}{N}\ln\sum_{s_t: m \mathrm{~fixed}} e^{- H_N}\equiv\Phi(m)=-\frac{ J}{2}m^2 - h_2(m) 
\end{equation}
where $h_2(m)$ is the binary entropy Bernoulli($\frac{1\pm m}{2}$). 
The van-der-Waals curve (or isotherm) is 
\begin{equation}\label{single-VdW}
h(m) = \frac{d \Phi(m)}{d m}=-Jm + \frac{1}{2}\ln\frac{1+m}{1-m},
\end{equation}
This relation can also be expressed as a fixed point equation for $m$ as a function of $h$, namely $m=\tanh(Jm+h)$, the CW equation.
Fig. \ref{vdw} shows $\Phi(m)$ and $h(m)$ for $J>1$ (dotted lines).
The van-der-Waals curve \eqref{single-VdW} depicted on figure \ref{vdw}
has two important thresholds: the static phase transition threshold at $h_c=0$ which satisfies the Maxwell equal area construction and the spinodal point
(or dynamic thresholds) which correspond to the points where the solutions of the fixed point equation bifurcate. 

One can consider a systems of CW models that are coupled along a chain: one constructs Hamiltonians where all spins at position 
$i$ are interacting with all spins at positions $[z-w, z+w]$. One can show that in the limit of an infinitely long chain,
when the window size grows the canonical free energy of the coupled chain becomes the convex hull of $\Phi(m)$. This phenomenon and the corresponding 
oscillations of the van-der-Waals curve around the Maxwell plateau are illustrated in figure \ref{vdw}. 

LDPC ensembles can be interpreted a spin models on random sparse graphs with interactions between the spins given by hard constraints. 
The code words are the allowed spin configurations. Their relative weight $\omega$ is a total magnetization $m$.
The number of code words  $A_n(\omega)$ of relative weight $\omega$ is 
a canonical partition function\footnote{let us point out that the weight enumerator polynomial is the 
grand-canonical partition function computed for a fixed magnetic field} computed at fixed magnetization; the growth rate 
is (minus) a canonical free energy; its derivative $h(\omega)$ is a van-der-Waals curve. The latter has two thresholds: a static phase transition one given by 
the Maxwell construction at $h_c$ and an iterative one $h_{\rm it}$ (the local max and min on fig. \ref{vdw}) given by the value of $h$ where the solutions of the first two fixed point equations
in \eqref{fixed-point} bifurcate. The growth rate of coupled LDPC ensembles behaves in the same way than the free energy of the CW chain.

\section{Analysis of the $(r,l, L)$ ensemble}\label{3}

For this ensemble the combinatorial and replica calculations yield the same set of equations. 
Here we sketch the main steps of the combinatorial method, following \cite{Miller-Burstein}. 
We recall that $(i, i+k)$ parametrizes edges connecting
variable nodes at $i$ with check nodes at $i+k$. We set $\omega^\prime = \frac{1-\omega}{2}$ and count the number of
codewords of total weight $W= \lfloor (2L+1)n\omega^\prime\rfloor$.
In the following expression the generating function variables $z^\prime_{i-k,i}$ 
attached to edges $(i-k, i)$  are set to $0$  for $i-k<-L$ and $i-k>+L$.
We have
\begin{multline*}
\mathbb{E}[A_n(\omega)]
=
\sum_{E_{i,j}}
\left(\prod_{i=-L}^L\prod_{k=0}^{l-1} \frac{E_{i,i+k}!(n-E_{i,i+k})!}{E!}\right)\\
\cdot\text{coeff}\left[\prod_{i=-L}^L \left(1+x \prod_{k=0}^{l-1}y^{\prime}_{i,i+k}\right)^{n},
x^W\prod_{i=-L}^L \prod_{k=0}^{l-1}y^{\prime\,E_{i,i+k}}_{i,i+k}\right]\\
\cdot\prod_{i=-L}^{L+l-1}\text{coeff}\Bigg[\left(\frac{\prod_{k=0}^{l-1} 
(1+z^{\prime}_{i-k,i})^{\frac{r}{l}}+\prod_{k=0}^{l-1}
(1-z^{\prime}_{i-k,i})^{\frac{r}{l}}}{2}\right)^{\frac{l}{r}n}\\ ,
\prod_{k=0}^{l-1}z^{\prime\,E_{i-k,i}}_{i-k,i}\Bigg].
\end{multline*}
To compute this expression in the large size limit one uses ($n\to +\infty$
\begin{equation*}
\lim_{n\to +\infty}\frac{1}{n}\text{coeff}\left[f(x)^n, x^{tn}\right] = \inf_x\left[\ln(f(x)) - t\ln x\right]
\end{equation*}
and performs a saddle point calculation of the sum. This yields
\begin{multline*}
G_{1,\mathrm{com}}(\omega)
=\lim_{L\to \infty}\frac{1}{2L+1}
\sup_{\alpha_{i,j}}\Bigg\{
-\left(\sum_{i=-L}^L\sum_{k=0}^{l-1}h_2(\alpha_{i,i+k})\right)\\
+ \inf_{x,y^\prime_{i,i+k}}\Bigg[\sum_{i=-L}^L 
\ln\left(1+x\prod_{k=0}^{l-1}y^\prime_{i,i+k}\right)-
(2L+1)\omega^\prime\ln x\\
-\sum_{i=-L}^L\sum_{k=0}^{l-1}\alpha_{i,i+k}\ln y^\prime_{i,i+k}\Bigg]\\
+\sum_{i=-L}^{L+l-1}\inf_{z^\prime_{i,k}}\Bigg[\frac{l}{r}
\ln\left(\frac{\prod_{k=0}^{l-1} (1+z^\prime_{i-k,i})^{\frac{r}{l}}+
\prod_{k=0}^{l-1}(1-z^\prime_{i-k,i})^{\frac{r}{l}}}2\right)\\
- \sum_{k=0}^{l-1}\alpha_{i-k,i}\ln z^\prime_{i-k,i}\Bigg]\Bigg\}.
\end{multline*}
The critical point equations are best written in terms of 
the new variables $h\equiv -\frac12 \ln x$, $\alpha^\prime_{i,j} \equiv 1- 2\alpha_{i,j}$, $z_{i,j} \equiv\frac{1-z^\prime_{i,j}}{1+z^\prime_{i,j}}$,
$y_{i,j} \equiv\frac{1-y^\prime_{i,j}}{1+y^\prime_{i,j}}$.
After a lengthy calculation the critical point equations give the generalization of \eqref{fixed-point} (these are also precisely the replica equations)
\begin{equation*}
\begin{cases}
y_{i,i+k} &= {z_{i,i+k}}^{\frac{r}{l}-1}\prod_{k\prime=0, k^\prime\ne k}^{l-1}{z_{i+k-k^\prime,i+k}}^{\frac{r}{l}}\\
z_{i-k,i} &=  \tanh\left(h + \sum_{k^\prime=0,k^\prime\ne k}^{l-1} \tanh^{-1}y_{i-k,i-k+k^\prime}\right)\\
\omega  &=\frac1{2L+1}\sum_{i=-L}^L\tanh\left(h + \sum_{k=0}^{l-1} \tanh^{-1}y_{i,i+k}\right)\\
\end{cases}
\end{equation*}
It is helpful to keep in mind the message passing interpretation of these equations: $y_{i,i+k}$ are the messages passed from check nodes at position $i+k$ into variable nodes at position $i$ and $z_{i-k,i}$ are the messages passed from variable nodes at position $i-k$ into checks
at position $i$. Note that the boundary condition has now become  $z_{i-k,i}=1$ for $i-k<-L$ and $i+k>L$.
Note also that $\alpha^\prime_{i,j}$ has expressed
in terms of other variables and therefore eliminated.
Once the non-trivial fixed point solution is computed for fixed 
$\omega$, replacing it into $G_{1,\mathrm{com}}$ one finds
\begin{multline*}
G_{1,\mathrm{com}}(\omega)
= \lim_{L\to \infty}
\frac1{2L+1}\Bigg[
\sum_{i=-L}^{L+l-1} \frac{l}{r} \ln\left(\frac{1+\prod_{k=0}^{l-1}{z_{i-k,i}}^{\frac{r}{l}}}2\right)\\
+\sum_{i=-L}^L \left[
\ln\left(\mathrm{e}^{h}\prod_{k=0}^{l-1}\left(1+y_{i,i+k}\right)+\mathrm{e}^{-h}\prod_{k=0}^{l-1}\left(1-y_{i,i+k}\right)\right)\right]\\
-\sum_{i=-L}^L\sum_{k=0}^{l-1}\ln\left(1+y_{i,i+k}z_{i,i+k}\right)
\Bigg]
-\omega h
\end{multline*}
This formula constitutes the basis for the numerical results 
described in section \ref{2}.  Again, we recognize in the first three terms the contributions of $(2L+1+l-1)n\frac{l}{r}$ checks, $(2L+1)n$ variable nodes and $(2L+1)nl$ edges.

\section{Analysis of the $(l,r,w,L)$ ensemble}\label{4}

In this case the combinatorial growth rate $G_{2,\mathrm{com}}(\omega)$ is strictly greater than $G(\omega)$, and its calculation 
will not be presented here due to lack of space. We use the replica method which leads to fixed point equations that are similar to
density evolution equations,
\begin{align*}
\widehat\Pi_i(y)=\frac{1}{w^{r-1}}&\sum_{k_1,\cdots,k_{r-1}=0}^{w-1}\int \prod_{j=1}^{r-1} dz_j \Pi_{i-k_j}(z_j)
\\  & \times
\delta\bigl(y-\prod_{j=1}^{r-1} z_j\bigr),
\\
\Pi_i(z) = \frac{1}{w^{l-1}}&\sum_{k_1,\cdots,k_{l-1}=0}^{w-1}\int \prod_{j=1}^{l-1} dy_j \widehat\Pi_{i+k_j}(y_j)
\\  & \times
\delta\bigl(z - \tanh(h+\sum_{j=1}^{l-1} \tanh^{-1} y_j\bigr),
\end{align*}
with the relation between $h$ and $\omega$,
\begin{align*}
\omega = \frac{1}{2L+1}
\sum_{i=-L}^{L}
\frac{1}{w^l}&\sum_{k_1,\cdots,k_{l}=0}^{w-1}
\int \prod_{j=1}^{l} dy_j \widehat\Pi_{i+k_j}(y_j)
\\  & \times
\tanh\bigl(h+\sum_{j=1}^{l} \tanh^{-1} y_j\bigr).
\end{align*}
The messages emanating from variable nodes at position $i$ and entering 
checks in the range $\{i,\cdots, i+w-1\}$ have density $\Pi_i(z)$. Those emanating from the checks at position $i$ and entering variables 
in the range $\{i-w+1,\cdots, i\}$ have density
$\widehat\Pi_i(y)$. Because of the randomized construction of the ensemble these densities are non trivial and the above equations do not reduce to 
equations for scalar variables. For the boundary conditions we have to set
$\Pi_i(z)= \delta(z-1)$ for $i\in \{-L-w+1,\cdots, -L-1\}$  and  $i\in \{L+1,\cdots,L+w-1\}$.

Once these equations are solved the most convenient way to obtain the growth rate is to use the integral relation 
\eqref{integral}.
Alternatively one can use the expression,
\begin{align*}
& G_2(\omega) = -\omega h + \lim_{L\to \infty} \frac{1}{2L+1}\sum_{i=-L}^{L}\biggl[
\\ &
\frac{l}{r}\frac{1}{w^r}
\sum_{k_1,\cdots,k_{r}=0}^{w-1}\int \prod_{j=1}^{r} dz_j \Pi_{i-k_j}(z_j)
\ln\biggl(\frac{1+\prod_{j=1}^{r} z_j}{2}\biggr)
\\ &
+ 
\frac{1}{w^{l}}\sum_{k_1,\cdots,k_{l}=0}^{w-1}\int \prod_{j=1}^{l} dx_j \widehat\Pi_{i+k_j}(y_j) 
\\ &
\times 
\ln\biggl(e^h\prod_{j=1}^{l}(1+y_j) + e^{-h}\prod_{j=1}^{l}
(1-y_j)\biggr)
\\ &
-
\frac{1}{w}\sum_{k=0}^{w-1}
\int \prod_{j=1}^{l} dy_j dz_j \widehat\Pi_{i+k}(y) \Pi_{i}(z)
\ln(1+y z)\biggr].
\end{align*}

A simplification of this formalism occurs in the limit of large 
$l$, $r$ with $R= 1- \frac{l}{r}$ fixed. Indeed the densities become Dirac distributions and the set of fixed point equations takes a form similar to those the density evolution equations of coupled LDPC 
ensemble for the binary erasure channel. A similar fact has been discussed in \cite{itw} for constraint satisfaction problems
such as $K\text{-SAT}$ and $Q$ coloring for large $K$ and $Q$. Details of this analysis will appear elsewhere. 


\section*{Acknowledgments}

{\small We are grateful to Ruediger Urbanke for instructive discussions and encouragement. 
The work of S. Hamed Hassani was supported by grant no 200021-121903 of the Swiss National Science Foundation. 
The work of Ryuhei Mori was supported by Grant-in-Aid for Scientific Research for JSPS Fellows (22$\cdot$5936), MEXT, Japan.}





\end{document}

%% file: definition.tex
\definecolor{TODO}{rgb}{0.6,0.6,0.6} 

\definecolor{TOCHECK}{rgb}{0.8,0.8,0.8} 

\newtheorem{theorem}{Theorem}

\newcommand{\btheo}{\begin{theorem}}
\newcommand{\etheo}{\end{theorem}}
\newcommand{\bproof}{\begin{proof}}
\newcommand{\eproof}{\end{proof}}
\newtheorem{definition}[theorem]{Definition}
\newcommand{\bdefi}{\begin{definition}}
\newcommand{\edefi}{\end{definition}}
\newtheorem{fact}[theorem]{Fact}
\newcommand{\bprop}{\begin{fact}}
\newcommand{\eprop}{\end{fact}}
\newtheorem{corollary}[theorem]{Corollary}
\newcommand{\bcor}{\begin{corollary}}
\newcommand{\ecor}{\end{corollary}}
\newtheorem{example}[theorem]{Example}
\newcommand{\bex}{\begin{example}}
\newcommand{\eex}{\end{example}}
\newtheorem{lemma}[theorem]{Lemma}
\newcommand{\blemma}{\begin{lemma}}
\newcommand{\elemma}{\end{lemma}}
\newtheorem{remark}[theorem]{Remark}
\newcommand{\bremark}{\begin{remark}}
\newcommand{\eremark}{\end{remark}}
\newtheorem{conj}[theorem]{Conjecture}
\newcommand{\bconj}{\begin{conj}}
\newcommand{\econj}{\end{conj}}

\def\0{{\tt 0}} 
\def\1{{\tt 1}} 
\def\?{{\tt *}} 

%% file: all-36.tex
\setlength{\unitlength}{1bp}%
\begin{picture}(120,150)(0,0)
\put(-60,0)
{
	\put(0,0){\includegraphics[ height=2in, width=1.7in]{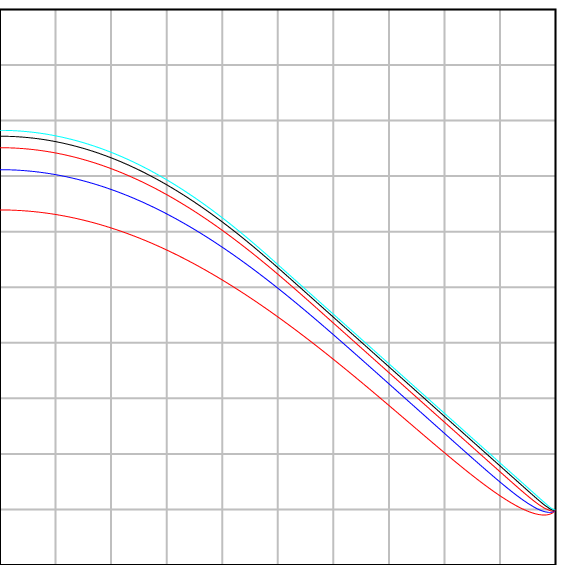}}
    \footnotesize	
        \put(2,-3){\makebox(0,0)[t]{$0$}}
	\put(121.5,-3){\makebox(0,0)[t]{$1$}}
        \put(-13,5){\makebox(0,0)[t]{$-0.05$}}
        \put(-5,17){\makebox(0,0)[t]{$0$}}
 	\put(-3,146){\makebox(0,0)[tr]{$0.5$}}

}
\put(66,0)
{
	\put(0,0){\includegraphics[ height=2in, width=1.7in]{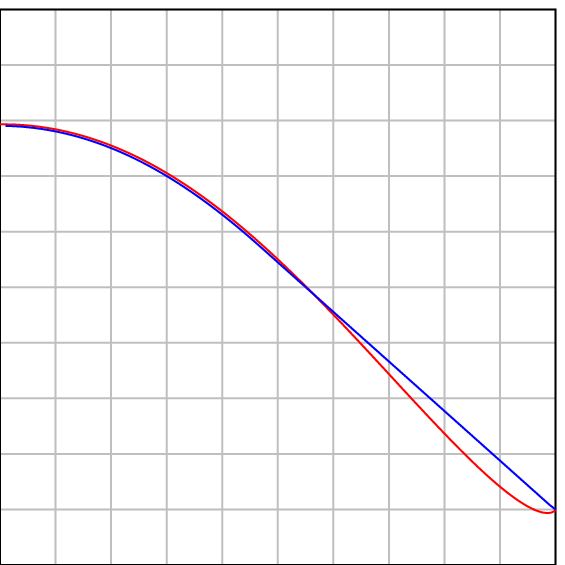}}
    \footnotesize
        \put(2,-3){\makebox(0,0)[t]{$0$}}
	\put(121.5,-3){\makebox(0,0)[t]{$1$}}	

}


\end{picture}

%% file: wiggle-we.tex
\setlength{\unitlength}{1bp}%
\begin{picture}(120,160)(0,0)
\put(-30,0)
{
	\put(0,0){\includegraphics[ height=2.4in, width=2.5in]{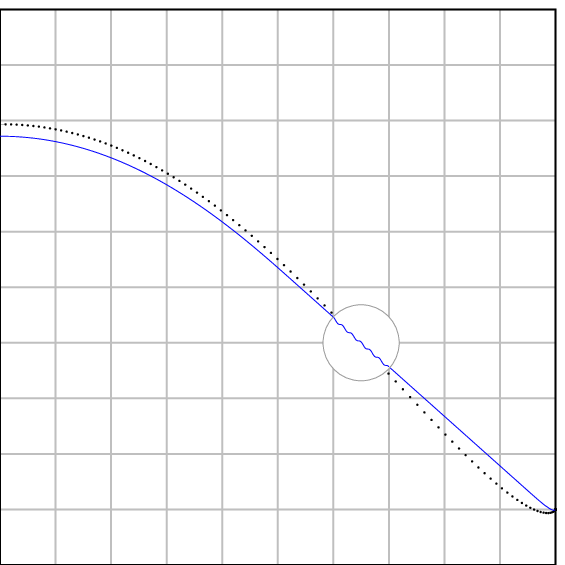}}
    \footnotesize	
 	\put(-3,6){\makebox(0,0)[tr]{$-0.05$}}
 	\put(-3,15){\makebox(0,0)[tr]{$0$}}
 	\put(-3,176){\makebox(0,0)[tr]{$0.45$}}
	\put(4,-3){\makebox(0,0)[tr]{$0$}}
 	\put(175,-3){\makebox(0,0)[tl]{$1$}}

}

\end{picture}

%% file: wiggle-hw.tex
\setlength{\unitlength}{1bp}%
\begin{picture}(120,160)(0,0)
\put(-30,0)
{
	\put(0,0){\includegraphics[ height=2.4in, width=2.5in]{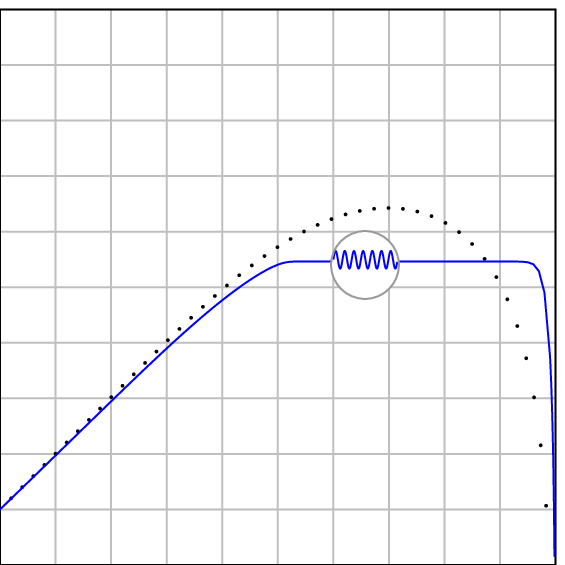}}
    \footnotesize	
 	\put(-3,6){\makebox(0,0)[tr]{$-0.1$}}
 	\put(-3,15){\makebox(0,0)[tr]{$0$}}
 	\put(-3,176){\makebox(0,0)[tr]{$0.9$}}
	\put(4,-3){\makebox(0,0)[tr]{$0$}}
 	\put(175,-3){\makebox(0,0)[tl]{$1$}}

}

\end{picture}

%% file: free-vdw.tex
\setlength{\unitlength}{1bp}%
\begin{picture}(120,110)(0,0)
\put(-65,0)
{
	\put(0,0){\includegraphics[height=1.5in, width=1.6in]{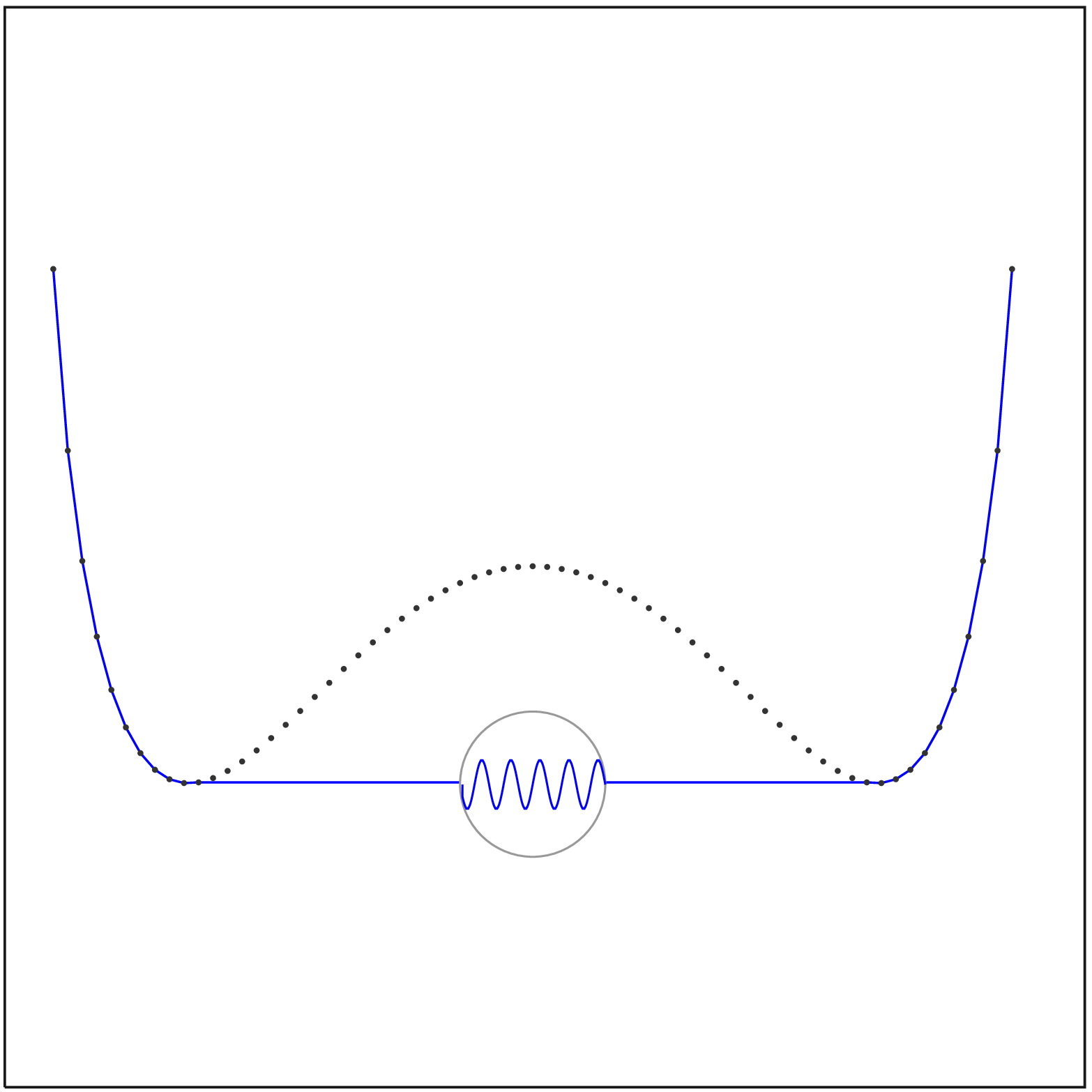}}
    \footnotesize	

}
\put(62,0)
{
	\put(0,0){\includegraphics[height=1.5in, width=1.6in]{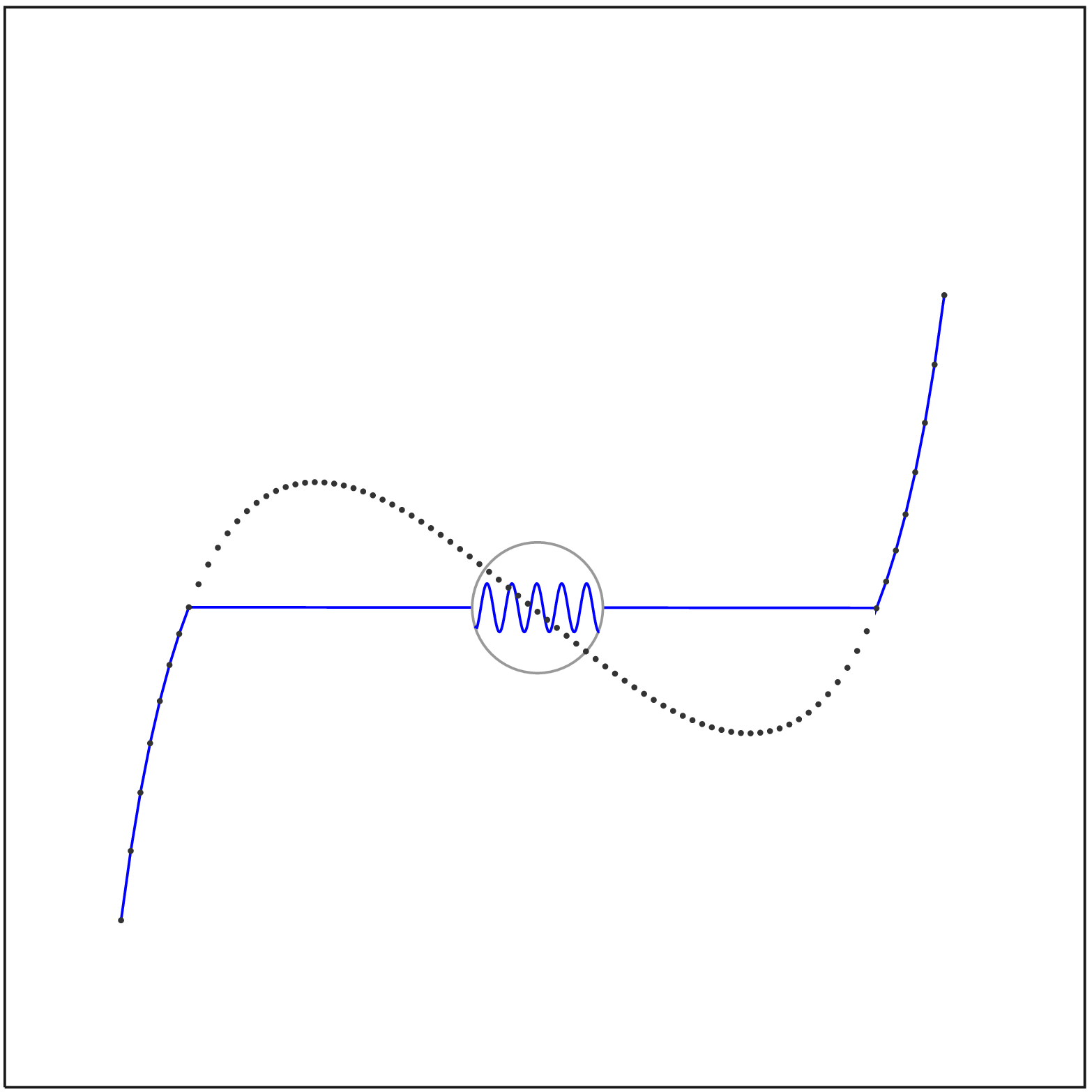}}
    \footnotesize	

}

\end{picture}